\documentclass[conference]{IEEEtran}
\IEEEoverridecommandlockouts
\usepackage{cite}
\usepackage{amsmath,amssymb,amsfonts}
\usepackage{algorithmic}
\usepackage{graphicx}
\usepackage{textcomp}
\usepackage{xcolor}
\usepackage{makecell}
\usepackage{multirow}

\usepackage{bbding}
\usepackage{color}
\usepackage{subfigure}
\usepackage{times}

\usepackage{soul}
\usepackage[hidelinks]{hyperref}
\usepackage[utf8]{inputenc}
\usepackage[small]{caption}
\usepackage{graphicx}
\usepackage{amsmath}
\usepackage{booktabs}
\usepackage{mathrsfs}
\usepackage{amsfonts,amssymb}
\usepackage{multirow}
\usepackage{color}
\usepackage{cite}
\usepackage{pifont}
\usepackage{array}
\usepackage[misc]{ifsym} 
\usepackage{makecell}
\usepackage{mathrsfs}
\usepackage{epstopdf}
\usepackage{threeparttable}
\def\BibTeX{{\rm B\kern-.05em{\sc i\kern-.025em b}\kern-.08em
T\kern-.1667em\lower.7ex\hbox{E}\kern-.125emX}}
\begin{document}

\title{  Histology Virtual Staining with Mask-Guided Adversarial Transfer Learning for Tertiary Lymphoid Structure Detection
\\
\thanks{This work was partially supported by the Independent Project of Yu-Yue Pathology Scientific Research Center (Grant No. YYKYXM202303B05) and the Chongqing Technical Innovation and Application Development Key Project (Grant No. cstc2021jscx-cylhX0005). 
Q.~Wang, Y.~Liu, and L.~Ma contributed equally to this paper.
Corresponding authors: W.~Chen, and X.~Yao.}
}

\author{
    \IEEEauthorblockN{
        Qiuli~Wang\textsuperscript{1,3},
        Yongxu~Liu\textsuperscript{1},
        Li~Ma\textsuperscript{2}, 
        Xianqi~Wang\textsuperscript{3},
        Wei~Chen\textsuperscript{3},
        and Xiaohong~Yao\textsuperscript{2}}
    \IEEEauthorblockA{\textsuperscript{1}Yu-Yue Pathology Research Center, Jinfeng Laboratory, Chongqing, 401329, China\\
    liu\_yongxu@outlook.com}
    \IEEEauthorblockA{\textsuperscript{2}Institute of Pathology and Southwest Cancer Center, \\Southwest Hospital, Army Medical University (Third Military Medical University), Chongqing, China\\
    mali@tmmu.edu.cn, yxh15@163.com}
    \IEEEauthorblockA{\textsuperscript{3}7T Magnetic Resonance Imaging Translational Medical Center, Department of Radiology,\\ Southwest Hospital, Army Medical University (Third Military Medical University), Chongqing, China\\
    \{wangqiuli, wxq9401, landcw\}@tmmu.edu.cn}
}

\maketitle

\begin{abstract}
Histological Tertiary Lymphoid Structures (TLSs) are increasingly recognized for their correlation with the efficacy of immunotherapy in various solid tumors. Traditionally, the identification and characterization of TLSs rely on immunohistochemistry (IHC) staining techniques, utilizing markers such as CD20 for B cells. Despite the specificity of IHC, Hematoxylin-Eosin (H\&E) staining offers a more accessible and cost-effective choice. Capitalizing on the prevalence of H\&E staining slides, we introduce a novel Mask-Guided Adversarial Transfer Learning method designed for virtual pathological staining. This method adeptly captures the nuanced color variations across diverse tissue types under various staining conditions, such as nucleus, red blood cells, positive reaction regions, without explicit label information, and adeptly synthesizes realistic IHC-like virtual staining patches, even replicating the positive reaction. Further, we propose the Virtual IHC Pathology Analysis Network (VIPA-Net), an integrated framework encompassing a Mask-Guided Transfer Module and an H\&E-Based Virtual Staining TLS Detection Module. VIPA-Net synergistically harnesses both H\&E staining slides and the synthesized virtual IHC patches to enhance the detection of TLSs within H\&E Whole Slide Images (WSIs). We evaluate the network with a comprehensive dataset comprising 1019 annotated slides from The Cancer Genome Atlas (TCGA). Experimental results compellingly illustrate that the VIPA-Net substantially elevates TLS detection accuracy, effectively circumventing the need for actual CD20 staining across the public dataset.
\end{abstract}

\begin{figure}[ht]
  \centerline{\includegraphics[width=0.4\textwidth]{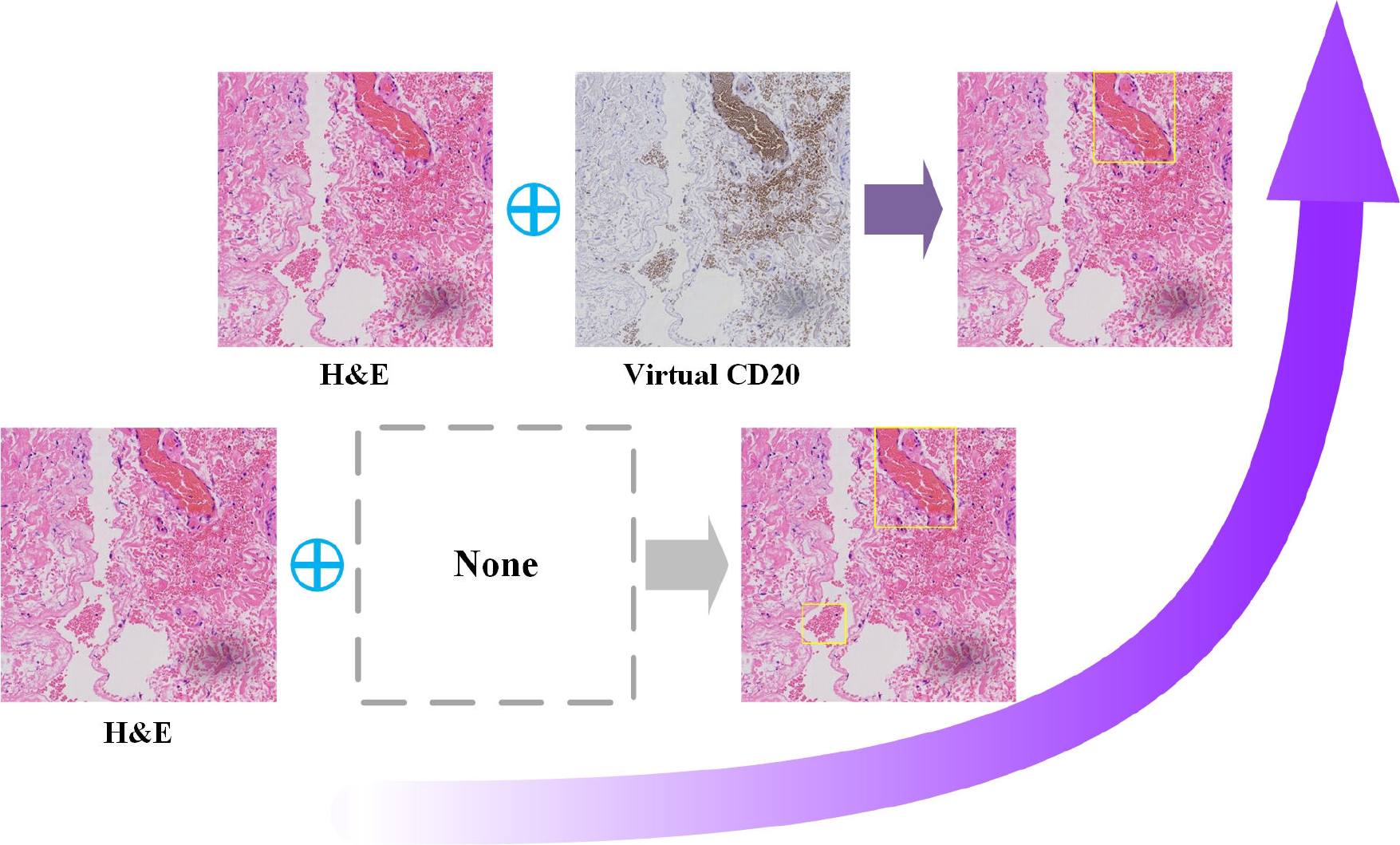}}
  \caption{H\&E Combined with CD20 Staining Improves The Detection Results. Our method significantly improves the detection of TLSs by accurately generating virtual CD20 staining patches, thereby reducing the false detection of TLSs.}
  \label{m0}
  \end{figure}

\IEEEpeerreviewmaketitle

\section{Introduction}
\IEEEPARstart{T}{ertiary} Lymphoid Structures (TLSs) in cancers are organized aggregates of immune cells that form postnatally immunohistochemistry tissues. TLSs are not found under physiological conditions but arise in the context of chronic inflammation \cite{schumacher_tertiary_2022}.
TLSs have been identified in several solid tumor types and are associated with better survival when present in the tumor microenvironment \cite{vanhersecke_mature_2021}.

Morphological analysis based on hematoxylin and eosin (H\&E) staining can achieve the quantification of TLSs as dense aggregates of lymphocytes, which have different contours compared to unorganized inflammation \cite{van_rijthoven_multi-resolution_2024}. The objective and reliable TLS analysis rely on immunohistochemistry (IHC) staining for various TLS-associated cell types such as B cells (CD20) or T cells (CD3, CD8) \cite{2014Presence,2008Long}, which is time-consuming and relies on the labor of pathologist.

Recently, several studies have proposed that synthesizing virtual IHC staining has the potential to decrease the demands for tissues and chemicals, lead to shorter times, and increase diagnosis efficiency \cite{latonen_virtual_2024,kapil_domain_2021}. Because of the low requirements for annotation and can be trained with unpaired data, CycleGAN-based methods \cite{kapil_domain_2021,8237506,Pradhan} have been a popular choice for virtual staining.

However, these methods suffer a general problem: Due to the lack of annotation, the synthesized pathological patches cannot reflect the tissues details of IHC staining and cannot synthesize reasonable IHC positive reaction.

To address this problem, we introduce a novel Mask-Guided Adversarial Transfer Learning method designed for virtual pathological staining. Based on CycleGAN, this method first captures pixel distributions from different pathological staining using the multi-Otsu threshold method, utilizing these distributions to automatically recognize tissue information such as aggregates of immune cells, red blood cells, \emph{etc}. Second, with this tissue information, the method constrains the synthesis of IHC staining details and generates reasonable IHC positive reactions, which provide extra information for TLS detection based on H\&E, as shown in Fig.~\ref{m0}.

Further, we propose the Virtual IHC Pathology Analysis Network (VIPA-Net), which contains a Mask-Guided Transfer Module and an H\&E-Based Virtual Staining TLS Detection Module. The Mask-Guided Transfer Module adopts the Mask-Guided Adversarial Transfer Learning method, taking H\&E patches as inputs and synthesizing target IHC patches. The Virtual Staining TLS Detection Module analyzes real H\&E patches and corresponding synthesized IHC patches to detect TLSs on H\&E Whole Slide Images (WSIs). 

The proposed method has two advantages:
(1) Low requirements for labels. The training of the synthesis module only needs unpaired H\&E and IHC WSI patches.
(2) Real-time adjustments for different datasets. The pixel distributions from different pathological staining can be re-calculated to better refine the network.

We evaluate our method on the dataset:
1019 manually annotated TCGA slides with over 10,000 TLS annotations, provided by the study \cite{van_rijthoven_multi-resolution_2024}, including lung squamous cell carcinoma (LUSC), clear cell renal cell carcinoma (KIRC), and muscle-invasive bladder cancer (BLCA).

The contributions of this study can be summarized as follows:
\begin{itemize}
\item We introduce a novel Mask-Guided Adversarial Transfer Learning method, which can generate reasonable IHC staining images under the control of unsupervised tissue information such as aggregates of immune cells, red blood cells, \emph{etc}.
\item We propose the Virtual IHC Pathology Analysis Network (VIPA-Net), which contains a Mask-Guided Transfer Module and an H\&E-Based Virtual Staining TLS Detection Module. The Virtual Staining TLS Detection Module analyzes real H\&E patches and the corresponding virtual IHC patches synthesized by the Mask-Guided Transfer Module to detect TLSs on H\&E Whole Slide Images (WSIs). 
\item We evaluate our method and experimental results demonstrate that our method can improve the TLS detection accuracy on the TCGA dataset.
\end{itemize}

\section{Related Work} 
\label{relatedwork}

\subsection{Traditional Methods for Detecting Tertiary Lymphoid Structures}
The common methodologies employed for the detection and quantification of Tertiary Lymphoid Structures (TLS) in the existing literature are diverse, encompassing techniques such as hematoxylin and eosin (H\&E) staining, multiplex immunohistochemistry (mIHC), multiplex immunofluorescence (mIF), and the real-time quantitative polymerase chain reaction (qPCR) of chemokines \cite{yang_detection_2023,helmink_b_2020,vanhersecke_mature_2021,van_dijk_preoperative_2020}. 

In certain prior studies, the identification of Tertiary Lymphoid Structures (TLSs) relied on multiplexed immunohistochemistry staining or multiplex immunofluorescence imaging \cite{helmink_b_2020,vanhersecke_mature_2021,cabrita_tertiary_2020,petitprez_b_2020,dieu-nosjean_quantitative_2018}. In 2018, Silia \cite{dieu-nosjean_quantitative_2018} systematically described the mIF staining technique and multispectral imaging procedure for TLSs. They introduced a quantitative pathology approach that employs a panel of seven markers, including CD21, CD23, CD20, CD3, DC-LAMP, PNAd, and DAPI, to assess TLSs across various stages of maturation. In 2020, Helmink \cite{helmink_b_2020} used both H\&E and CD20 immunohistochemistry staining to qualify and quantify TLSs. Their analysis showed that CD20+ B cells were localized in TLSs, and were colocalized with CD4+, CD8+, and FOXP3+ T cells, as well as CD21+ follicular dendritic cells and MECA79+ high endothelial venules. In 2021, Vanhersecke \cite{vanhersecke_mature_2021} screened for TLSs in tumors with H\&E and CD3/CD20 immunohistochemistry and evaluated their maturity using a multiplex immunofluorescence assay combining CD4, CD8, CD20, CD21 and CD23 markers. They also demonstrated that the dual CD20 and CD23 immunostaining approach yielded efficacy akin to multiplex immunofluorescence in evaluating the maturity of tertiary lymphoid structures.

\begin{figure*}[ht]
  \centerline{\includegraphics[width=1\textwidth]{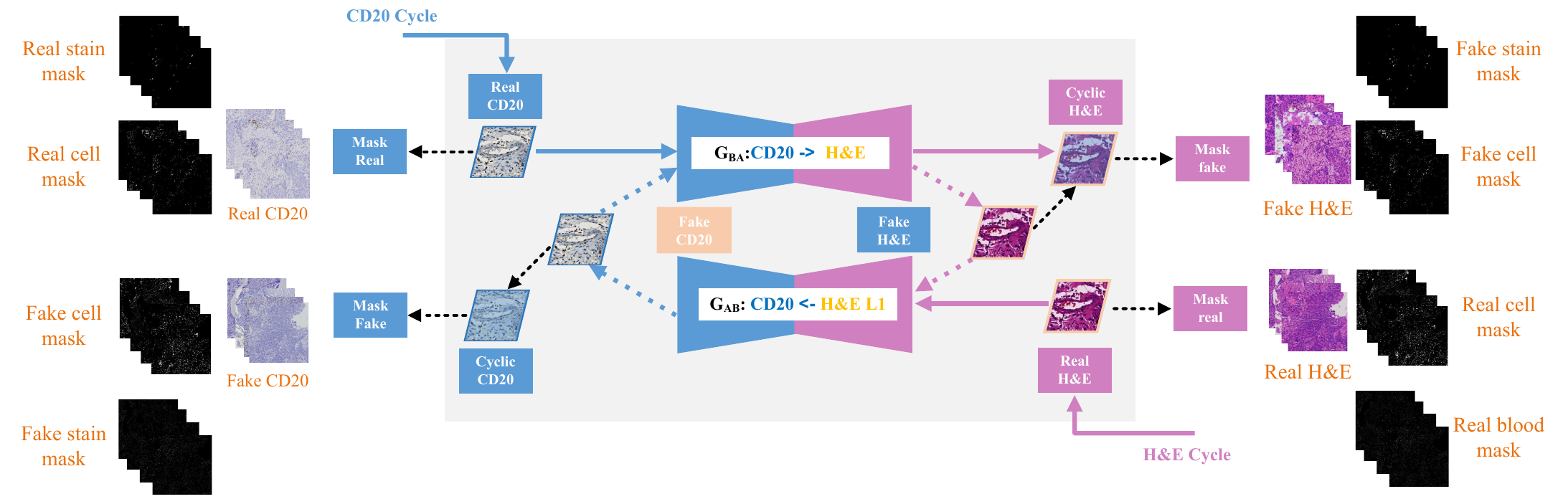}}
  \caption{Mask-Guided Adversarial Transfer Learning. In this module, the left side contains the real CD20 staining patches, the virtual CD20 staining patches, and the corresponding automatically generated cell and stain masks. And the right side contains the real H\&E staining patches, the virtual H\&E staining patches, and the corresponding automatically generated nucleus and red blood cell masks.}
  \label{VIPA-Net}
  \end{figure*}

These studies presented a feasible strategy for the identification of TLSs, where IHC staining effectively improves detection outcomes. However, the substantial cost and complexity of CD20 staining limit its further use. Therefore, we aim to propose an alternative to the current multiplex imaging-based approach for identifying TLSs that is both widely available and practical. Hematoxylin and eosin (H\&E) staining, commonly used in clinical practice, is also applicable for detecting tertiary lymphoid structures (TLS) in tumors. Several groups have evaluated TLSs in H\&E staining slides through pathologist evaluation \cite{calderaro_intra-tumoral_2019} and have highlighted that this approach is time-consuming and labor-intensive, as well as prone to interobserver variability \cite{buisseret_reliability_2017}. Given the emphasis in prior studies on the significance of CD20 for the detection of tertiary lymphoid structures (TLS), we intend to apply virtual staining techniques to commonly available H\&E staining slides to create CD20 staining immunohistochemical images. Unlike previous methods that rely on extensive manual annotation for fully supervised training, we use thresholding to automatically generate mask annotations, enabling self-supervised style transfer training. This approach integrates the precision of immunohistochemistry with the convenience of H\&E staining for TLS detection.

\subsection{Artificial Intelligence Approaches for Tertiary Lymphoid Structures  Detection}
Artificial Intelligence is increasingly used to extract clinically relevant information from digital pathology data \cite{xu_whole-slide_2024,lu_data-efficient_2021,coudray_classification_2018}. Various studies demonstrated the practicality of deep learning for automated diagnosis, classification, genetic alteration prediction, and prognostic assessment of cancer using Hematoxylin and Eosin (H\&E) images \cite{lu_prognostic_2020,lu_ai-based_2021,strom_artificial_2020,bulten_automated_2020,kather_deep_2019,skrede_deep_2020,yamashita_deep_2021,bilal_development_2021,muti_development_2021,fu_pan-cancer_2020,kather_pan-cancer_2020,mobadersany_predicting_2018}.

Furthermore, several deep learning models have been developed towards automated segmentation of TLSs from H\&E images in various tumor types \cite{10349929,wang_computerized_2023,chen_deep_2024,li_development_2023,van_rijthoven_hooknet_2021,barmpoutis_tertiary_2021,van_rijthoven_multi-resolution_2024}. Barmpoutis \cite{barmpoutis_tertiary_2021} identified regions of Tertiary Lymphoid Structures (TLS) in lung cancer tissues by applying a method that combines a DeepLab v3+ network, an active contour model, and a lymphocyte segmentation approach. Van Rijthoven \cite{van_rijthoven_hooknet_2021} developed a multiresolution convolutional neural network known as HookNet, which can automatically and more accurately identify mature Tertiary Lymphoid Structures (TLSs) with Germinal Centers (GCs) compared to previous TLS identification methods. Li Z \cite{li_development_2023} constructed a deep learning–based single-cell analysis tool capable of performing automated segmentation and classification of tumor-infiltrating lymphocytes on whole-slide images for automated detection, enumeration, and classification of TLSs in H\&E staining WSIs. Wang Y \cite{wang_computerized_2023} devised an automated computational pipeline for quantifying the TLSs in the tumor region of routine H\&E staining WSIs. The research by Chen Z \cite{chen_deep_2024} differs from studies that rely solely on pathologists' manual annotation of TLS without the guidance of multiplex immunohistochemistry (mIHC). They utilized mIHC markers-DAPI, CD3, and CD20 to identify TLS, thereby reducing the impact of human subjective judgment. Wang B \cite{10349929} proposed a weakly supervised segmentation network that incorporates cross-scale attention guidance and noise-sensitive constraint for TLS detection. 

Taking into account the guiding role of immunohistochemistry in the detection of TLS, we conduct virtual immunohistochemical staining on H\&E staining WSIs to simulate the actual detection process of TLS.

\section{Virtual IHC Pathology Analysis Network}
\label{methods}
The proposed Mask-Guided Adversarial Transfer Learning method can improve the virtual staining in two ways:
lower demands for labels, and better details for different staining.
To apply this method, we propose a Virtual IHC Pathology Analysis Network (VIPA-Net), which contains a Mask-Guided Transfer Module and an H\&E-Based Virtual Staining TLS Detection Module, as shown in Fig.~\ref{VIPA-Net}. The Virtual Staining TLS Detection Module analyzes real H\&E patches and corresponding virtual IHC patches synthesized by Mask-Guided Transfer Module to detect TLSs on H\&E Whole Slide Images (WSIs).

\subsection{Mask-Guided Adversarial Transfer Module}
This module can transfer H\&E patches to realistic IHC patches, and conversely transfer IHC patches to H\&E patches. 
The backbone of this module is CycleGAN, a widely used deep learning model for domain adaptation, which can be modified into an end-to-end network for joint staining adaptation \cite{kapil_domain_2021}.

\subsubsection{Unsupervised Mask Extraction}
\label{mask_extraction}
\begin{figure}[htbp]
  \centerline{\includegraphics[width=0.5\textwidth]{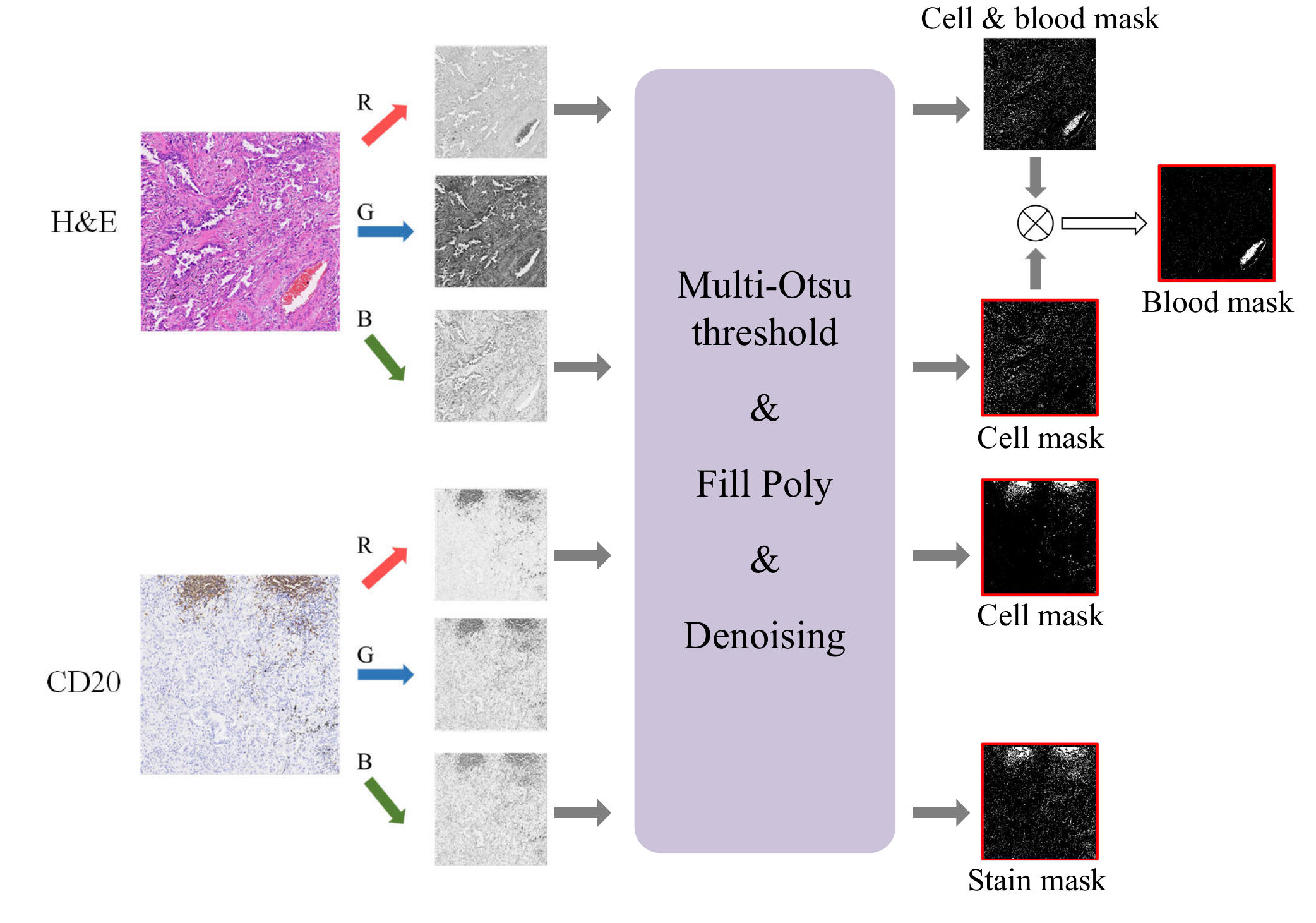}}
  \caption{Unsupervised Mask Extraction. Extract the best-performing mask images from different channels of the staining patches. This process uses thresholds to automatically extract the masks, avoiding the time-consuming and labor-intensive manual annotation.}
  \label{mask_process}
  \end{figure}

Pathological staining colors different tissues in distinct ways, leading to different pixel distributions in the RGB channels \cite{10349929}. In this study, we further propose that by analyzing pixel distributions in the three channels with threshold method, we can extract coarse but effective masks that indicate nucleus spacial and tissue information unsupervised, which is shown in Fig.~\ref{mask_process}.

The unsupervised mask extraction process can be summarized as follows:
(1) Extract patches from WSI images using code provided by M. Y. Lu \cite{lu_data-efficient_2021}. 
(2) Split each patch $x$ into Red $x_r$, Green $x_g$, Blue $x_b$ three greyscale channels. 
(3) Extract target regions with Otsu in selected channel according to staining style. We set 7 Otsu thresholds to divide WSI patches into 8 groups, and then use the second highest threshold to extract unsupervised masks.
(4) Remove noise pixels and fill holes in target regions.

Please note that step (3) should be adjusted according to pathological staining type.
For H\&E staining WSIs, we extract nucleus $m_{n}$ in $x_b$, then extract nucleus and red blood cells $m_{n+r}$ together in $x_r$. we have mask for nucleus $m_{n}$ and mask for red blood cells $m_r = m_{n+r}\oplus m_n$. 
For CD20 WSIs, we extract nucleus $m_{n}$ in $x_b$, then extract IHC positive mask $m_p$ in  $x_g$.

\begin{table*}[htbp]
  \centering
  % \resizebox{\textwidth}{12mm}
  \caption{The Number of Dataset Components.}
  \scalebox{1}{
  \label{b1}
  % \begin{tabular}{cccccccc}
  % \begin{tabular}{p{4cm}p{1cm}p{1cm}p{2cm}p{2cm}p{1cm}p{2cm}p{2cm}}
  \begin{tabular}{>{\centering\arraybackslash}m{4cm} >{\centering\arraybackslash}m{1cm} >{\centering\arraybackslash}m{1cm} >{\centering\arraybackslash}m{1.5cm} >{\centering\arraybackslash}m{2cm} | >{\centering\arraybackslash}m{1cm} >{\centering\arraybackslash}m{2cm} >{\centering\arraybackslash}m{1.5cm}}

  \toprule   %\multicolumn{2}{|c|}{合并一行两列}
    \multirow{2}{*}{\centering \textbf{The name of dataset}} & \multicolumn{4}{c}{Train} & \multicolumn{3}{c}{Val} \\ \cmidrule{2-8}
     & \textbf{H\&E} & \textbf{CD20} & \textbf{H\&E\_patches} & \textbf{CD20\_patches} & \textbf{H\&E} & \textbf{H\&E\_patches non-overlapping} & \textbf{H\&E\_patches overlapping} \\ \midrule
    The TCGA TLS pathology dataset & 32 & 38 & 6000 & 4779 & 8 & 893 & 893 \\ 
    % The TCGA TLS pathology dataset & - & - & - & - & 1018 & 47469 & 72548 \\
  \bottomrule
  \end{tabular}
  }
  \begin{tablenotes} 
  \item The TCGA dataset includes 6,000 H\&E staining patches and 4,779 CD20 staining patches in the training phase, and 893 H\&E staining patches in the validation phase. 
  % There are 47,469 H\&E staining patches in the TCGA dataset without overlapping segmentation, and 72,548 H\&E staining patches in the TCGA dataset with overlapping segmentation. 
  \end{tablenotes} 
  \end{table*}

  % \begin{table*}[htbp]
  %   \centering
  %   % \resizebox{\textwidth}{12mm}
  %   \caption{The Number of Datasets Components.}
  %   \scalebox{1}{
  %   \label{b1}
  %   \begin{tabular}{ccccc}
  %   \toprule   %\multicolumn{2}{|c|}{合并一行两列}
  %    & \multicolumn{2}{c}{Train} & \multicolumn{2}{c}{Val} \\ \midrule
  %   \textbf{The name of datasets} & \textbf{H\&E} & \textbf{CD20} & \textbf{H\&E} & \textbf{CD20} \\ \midrule
  %     The PRIVATE TLS pathology dataset & 6000 & 4779 & 893 & - \\ 
  %     The TCGA TLS pathology dataset & - & - & 47469 / 72548 & - \\
  %   \bottomrule
  %   \end{tabular}
  %   }
  %   \begin{tablenotes} 
  %   \item We only use the PRIVATE dataset to train the generation model, so the TCGA dataset does not include the training part. The PRIVATE dataset includes 6,000 H\&E staining patches and 4,779 CD20 staining patches in the training phase, and 893 H\&E staining patches in the validation phase. There are 47,469 H\&E staining patches in the TCGA dataset without overlapping segmentation, and 72,548 H\&E staining patches in the TCGA dataset with overlapping segmentation. 
  %   \end{tablenotes} 
  %   \end{table*}

\subsubsection{Mask-Guided Adversarial Transfer Learning}
To provide accurate vessel synthesis in H\&E and positive reaction synthesis in CD20, we propose the Mask-Guided Adversarial Transfer Learning, which utilizes unsupervised masks of different pathological staining WSIs as the guidance of virtual transfer learning.
Unlike conventional CycleGAN, the proposed Mask-Guided Adversarial Transfer Module uses masks that indicate specific regions to guide the WSI synthesis.

Since this module is based on CycleGAN, it involves two cycles: $G_{AB} \rightarrow G_{BA}$ and $G_{BA} \rightarrow G_{AB}$.
Two generators $G_{AB}:x_a \rightarrow x_b$ and $G_{BA}:x_b \rightarrow x_a$, where $x_a\in x_A$ and $x_b\in x_B$, are trained to synthesize samples in domain A (H\&E) in to domain B (IHC) and vice versa. Parameters in this module are learned with two adversarial losses $\mathcal{L}^{AB}_{GAN}$ and $\mathcal{L}^{BA}_{GAN}$:
\begin{align}
  \mathop{\min}_{G_{AB}}\mathop{\max}_{D_B} \mathcal{L}^{AB}_{GAN} = \mathbb{E}_{x_B} \sim x_B log(D_B (x_B)) + 
  \notag
  \\ \mathbb{E}_{x_A} \sim x_A log(1-D_B (G_{AB}(x_A)))
  \\ 
  \notag
  \\
  \mathop{\min}_{G_{BA}}\mathop{\max}_{D_A} \mathcal{L}^{BA}_{GAN} = \mathbb{E}_{x_A} \sim x_A log(D_A (x_A)) + 
  \notag
  \\ \mathbb{E}_{x_B} \sim x_B log(1-D_A (G_{BA}(x_B)))
\end{align}
where two discriminators $D_A$ and $D_B$ are trained in opposition to identify synthetic from real samples in the two domains.

Conventional methods use a cycle-consistent loss to prevent mode collapse of two adversarial modules \cite{8237506}.
Unlike conventional methods, we modify the conventional cycle consistent loss $\mathcal{L}_{cycle}$ and use masks captured by unsupervised threshold method to better guide the synthesis. The $\mathcal{L}_{cycle}$ is defined as:
\begin{align}
  \mathbb{E}_{x_A} \sim x_A \Vert x_A-G_{BA}(x'_B) \Vert_1 + \\
  \notag
  \mathbb{E}_{m_A} \sim m_A \Vert m_A-m_B \Vert_1 + \\
  \notag
  \mathbb{E}_{x_B} \sim x_B \Vert x_B-G_{AB}(x'_A) \Vert_1 + \\
  \notag
  \mathbb{E}_{m_B} \sim m_B \Vert m_B-m_A \Vert_1
\end{align}
where $m$ indicates masks extracted by unsupervised method.

Generally speaking, $G_{AB} \rightarrow G_{BA}$ takes a real H\&E patch $x_{he}$ as the input, transfers $x_{he}$ to virtual CD20 patch $x'_{cd20}$, and finally transfer $x'_{cd20}$ to virtual H\&E patch $x'_{he}$. 
As a result, the first cycle can provide a synthesis $x'_{he}$ and a corresponding real $x_{he}$. If two generators can accurately synthesis pathological staining details, $x'_{he}$ and $x_{he}$ should have similar masks under the same thresholds.

Similarly, $G_{BA} \rightarrow G_{AB}$ takes a real CD20 patch $x_{cd20}$ as input, and provides a synthesis $x'_{cd20}$. $x'_{cd20}$ and $x_{cd20}$ should have similar masks under the same thresholds.

\subsection{H\&E-Based Virtual Staining TLS Detection Module}
\begin{figure}[ht]
\centerline{\includegraphics[width=0.5\textwidth]{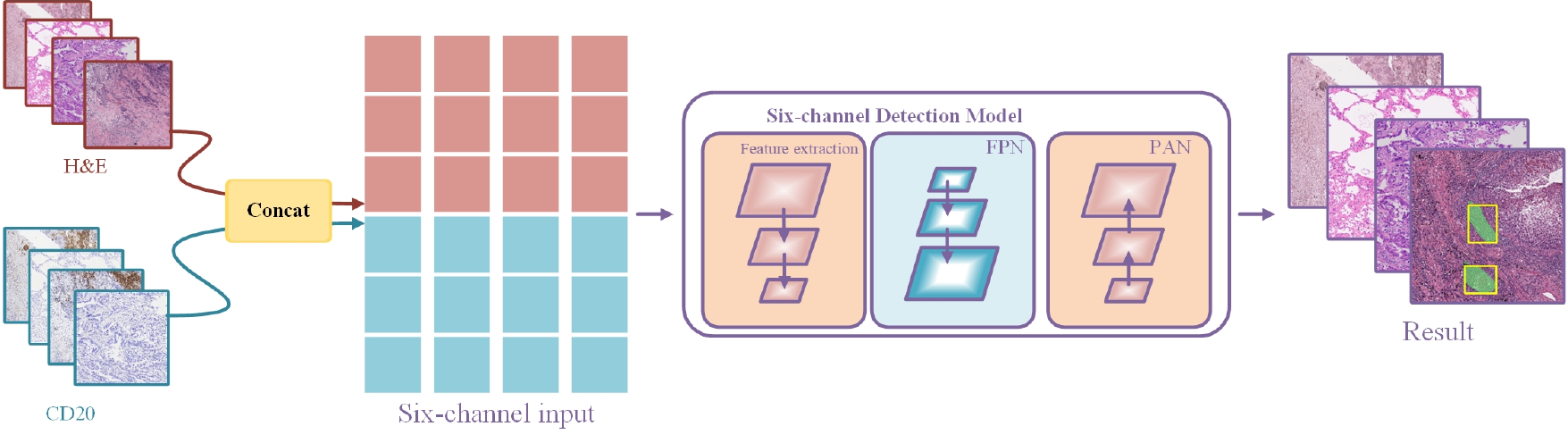}}
\caption{The Detection Models input concatenation Process. We concatenate the RGB three-channel patches of H\&E and CD20 into six-channel patches and then input them into the detection model to obtain their respective predicted results, which are shown with yellow candidate boxes.}
\label{m2}
\end{figure}

The input of this module contains a real H\&E patch $x_{he}$ and a corresponding synthesized CD20 patch  $x'_{cd20}$. As shown in Fig.~\ref{m2}. The backbone of this module contains a six-channel input of YOLOv5, which has been proved as an effective detection model. The H\&E and CD20 staining patches are concatenated one-to-one, forming six-channel data that is input into the model.

% \begin{figure}[ht]
%   \centerline{\includegraphics[width=0.4\textwidth]{m2.pdf}}
%   \caption{The Detection Models Result Merged Process. H\&E and CD20 staining patches are input into two specialized detection models to obtain their respective predicted results, which are shown with yellow candidate boxes. The final results are obtained through non-maximum suppression (NMS), which combines the predicted boxes sorted according to the confidence level. }
%   \label{m2}
%   \end{figure}

\begin{figure*}[htbp]
  \centerline{\includegraphics[width=0.95\textwidth]{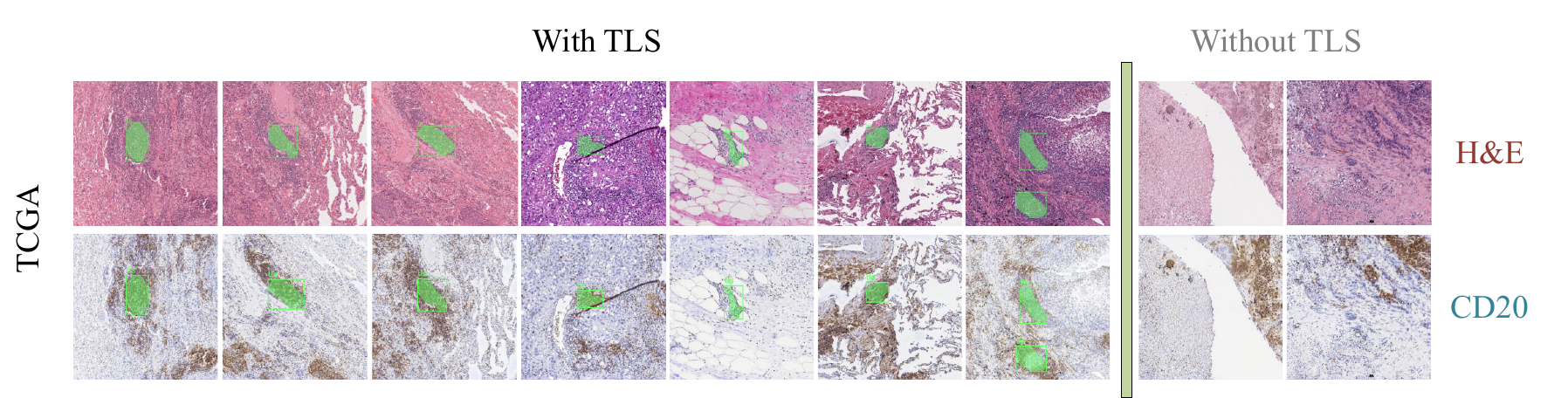}}
  \caption{The Dataset Visualization. For the TCGA dataset, we compare the effect of H\&E staining patches and mask-guided CD20 staining patches. In the images, the left side of the green dividing line shows patches with the presence of TLS, while the right side displays patches without TLS. The green mask represents the ground truth of TLS.}
  \label{m1}
  \end{figure*}

\section{Experiments}
\label{experiments}
\subsection{Datasets and Experimental Settings}
\textbf{Experimental setup. }We evaluate our method using a public dataset: a public TLS pathology dataset from TCGA. For the public TCGA dataset, which includes H\&E and CD20 staining patches, we divide it into an 8:2 ratio for training and validation. In this dataset, variations in staining time can cause slight differences in staining, potentially affecting the models training. Therefore, we employ staining normalization to mitigate this effect. The quality of the processed dataset is then tested to ensure the reliability of the results obtained by the models. We use two NVIDIA RTX A6000 GPUs for training both the generative model and the detection model. The patch size is 512 pixels, and we train for 100 epochs. All other model parameters are set to their default values.

\textbf{TCGA TLS pathology dataset. }We download 1,019 manually annotated TCGA slides with over 10,000 TLS annotations. All analyzed H\&E images are scanned at 40$\times$ magnification in Aperio SVS format. An initial manual quality control is performed, and the following exclusion criteria for images are applied: poor picture quality, insufficient information, and rare issues. Then, we use CLAM \cite{lu_data-efficient_2021} to crop the full WSI images at level 1, ensuring each patch is 512 pixels in size. Consequently, we use this TCGA H\&E and CD20 staining dataset for training and validating the proposed generative model. The dataset contains 6,893 H\&E staining patches and 4,779 CD20 staining patches, with an 8:2 ratio for training and validation

\begin{figure}[htp]
  \centerline{\includegraphics[width=0.4\textwidth]{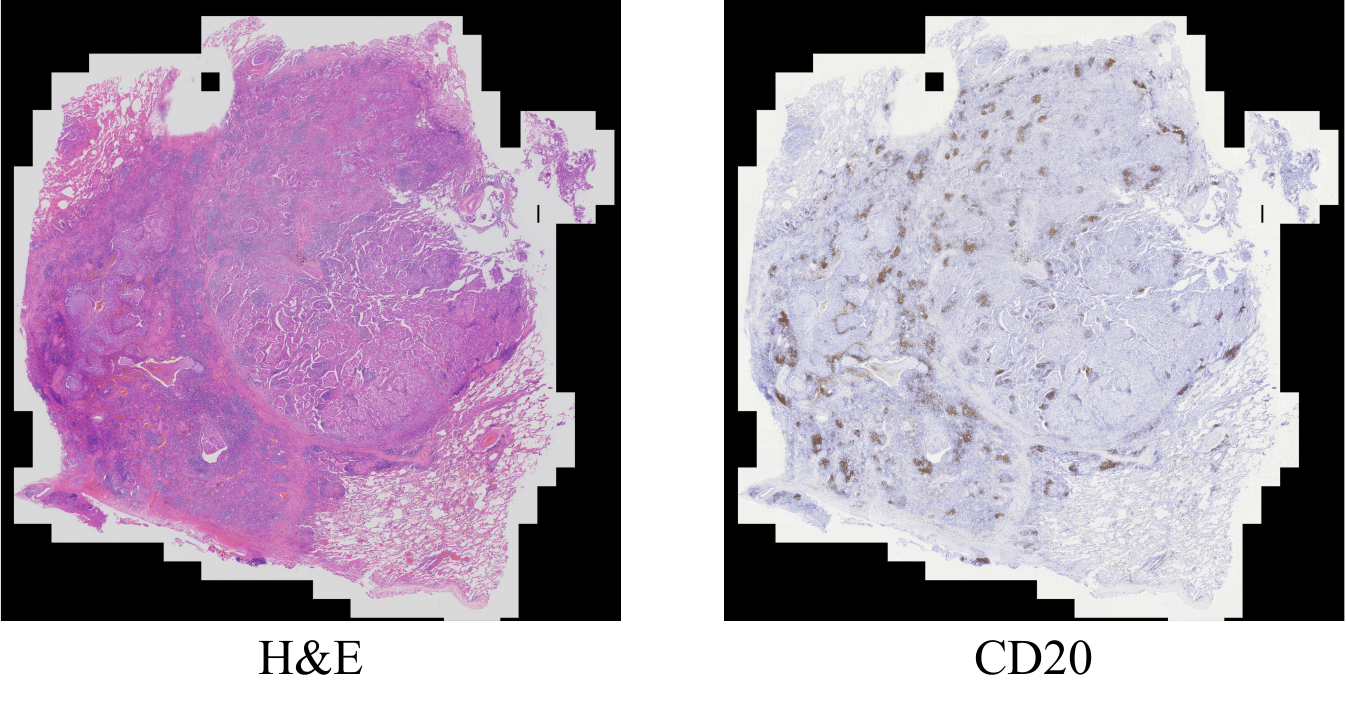}}
  \caption{Comparison of The WSIs. All generated CD20 staining patches are stitched together according to their coordinates. The left side shows the result of stitching H\&E patches, while the right side shows the result from the mask-guided CycleGAN model. The mask-guided CycleGAN model accurately generates colors in the red blood cells regions, outperforming the standard CycleGAN in patch generation }
  \label{mpj}
  \end{figure}
    
  \begin{figure*}[htp]
    \centerline{\includegraphics[width=1\textwidth]{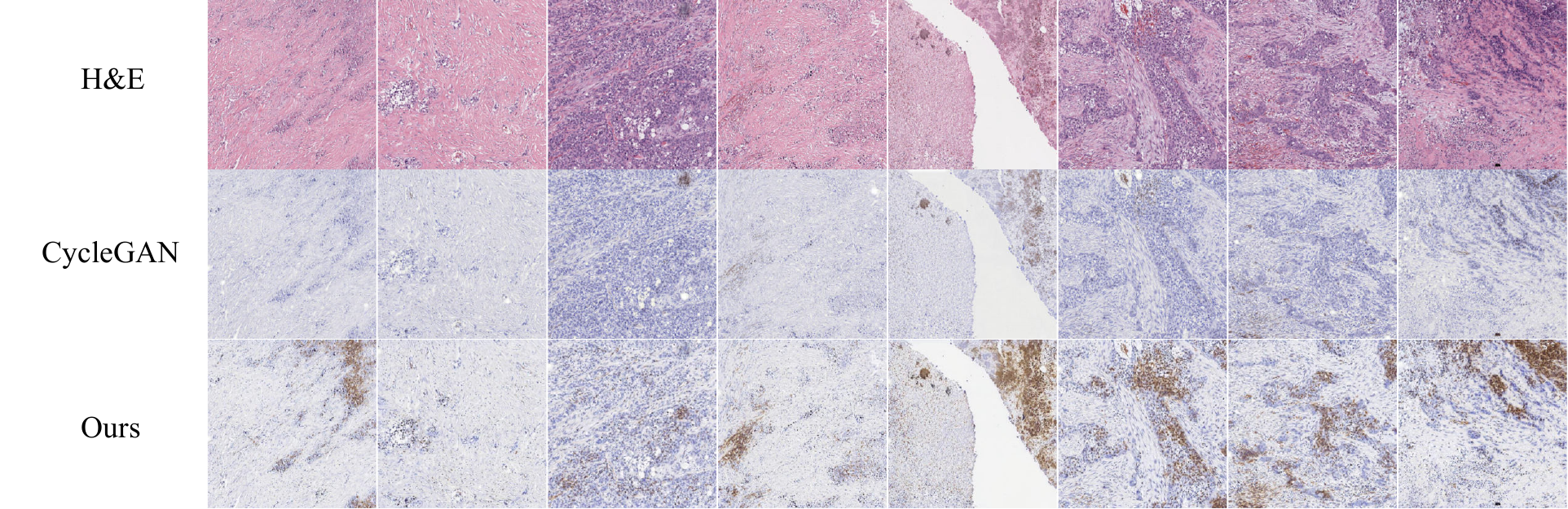}}
    \caption{The Comparison of Different Generative Models' Result. Each row in the image, from top to bottom, represents H\&E staining patches, CycleGAN staining patches, and mask-guided CycleGAN staining patches, respectively. As can be observed, mask-guided CycleGAN is more realistic in terms of generation.}
    \label{m3}
    \end{figure*}
  
  \begin{figure*}[htp]
    \centerline{\includegraphics[width=1\textwidth]{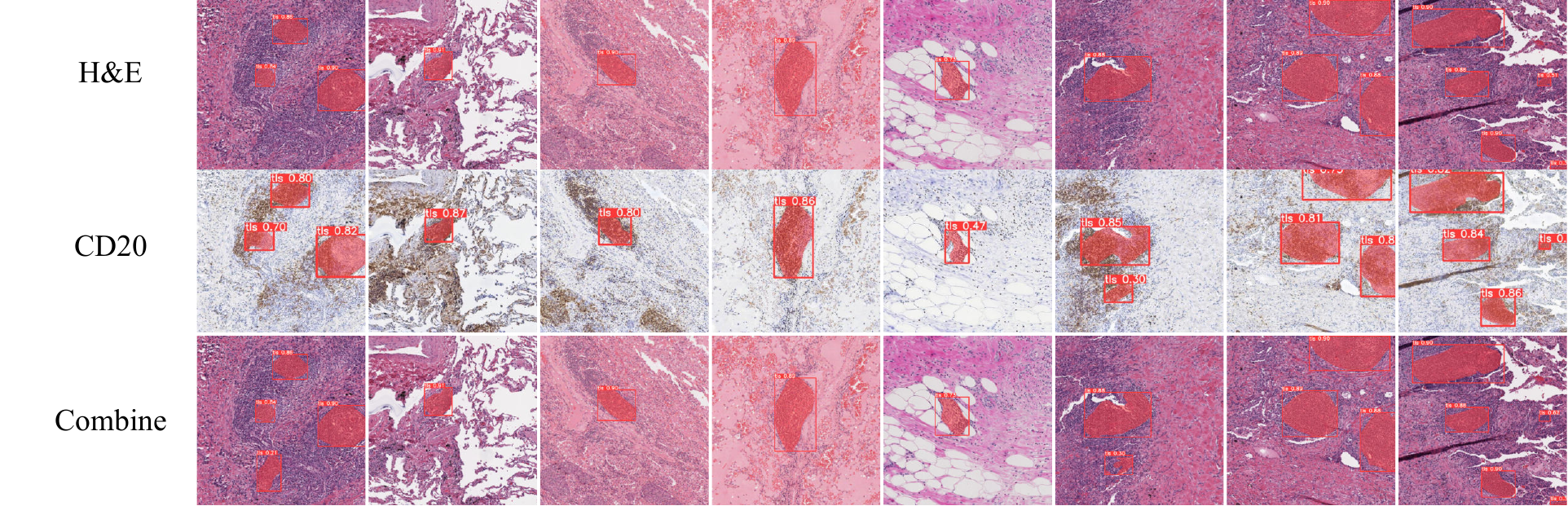}}
    \caption{Visualization of The Detection Models' Result. H\&E segmentation is better, while CD20 can detect TLS missed by H\&E. By combining H\&E and CD20 staining patches, we can get better detection results.}
    \label{m4}
    \end{figure*}

\subsection{Model Training and Validation}
Considering data heterogeneity, differences in staining results due to staining methods, long-tailed recognition, and poor results of H\&E detection due to its inherent structure, these factors pose a unique challenge to the precise localization of TLS. Therefore, we hope to generate more realistic CD20 staining patches by introducing a mask-guided generative model for H\&E patches. For the generative model, we use both H\&E and CD20 datasets (including masks generated by self-supervised methods for both datasets) to train it. The visualization of the datasets and their respective component counts are illustrated in Fig.~\ref{m1} and Table~\ref{b1}. For the detection model, we use the two staining datasets to train two separate detection models and then merge the detection results by non-maximum suppression (NMS) at the final evaluation, as shown in Fig.~\ref{m2}. For all model training and inference processes, we ensure that the original patch size is 1024 pixels. However, due to computing device limitations and the need to reduce the time required for training, we uniformly scale the patch size to 512 pixels. The CD20 staining patches of TCGA dataset are generated virtually.

\begin{table}[htbp]
  \centering
  % \resizebox{\textwidth}{12mm}
  \caption{The Validation Index of Detection Models.}
  % \scalebox{0.2}{
  \label{bfid}
  \begin{tabular}{ccccccc}
  \toprule   %\multicolumn{2}{|c|}{合并一行两列}
    %  & \multicolumn{2}{c}{train} & \multicolumn{2}{c}{val} \\ \midrule
  \textbf{The Name of Generative Model} & \textbf{FID} \\ \midrule
    CycleGAN & 43.195 \\ 
    Mask-guided CycleGAN & 33.525 \\
  \bottomrule
  \end{tabular}
  % }
  \begin{tablenotes} 
    \item We calculate the similarity of the generated pictures using 893 validation patches in the TCGA dataset.
  \end{tablenotes} 
  \end{table}
  
  \begin{table*}[htbp]
    \centering
    % \resizebox{\textwidth}{12mm}
    \caption{The Validation Index of Detection Models.}
    % \scalebox{0.2}{
    \label{b2}
    \begin{tabular}{ccccccc}
    \toprule   %\multicolumn{2}{|c|}{合并一行两列}
    %  & \multicolumn{2}{c}{train} & \multicolumn{2}{c}{val} \\ \midrule
    \textbf{The Name of Detection Model} & \textbf{P\_box} & \textbf{R\_box} & \textbf{F1\_box} & \textbf{P\_mask} & \textbf{R\_mask} \\ \midrule
      Only training by H\&E datasets & 86.4 & 80.4 & 83.29 & 86.6 & 80.2 \\ 
      Only training by CD20 datasets & 86.0 & 78.1 & 81.86 & 86.6 & 77.6 \\
      Combine       & 89.5 & 81.9 & 85.53 & 89.8 & 81.6 \\
    \bottomrule
    \end{tabular}
    \begin{tablenotes} 
      \item Among them, \textbf{P\_box}, \textbf{R\_box}, and \textbf{F1\_box} represent the precision, recall, and F1-score of the predicted bounding boxes, respectively. \textbf{P\_mask} and \textbf{R\_mask} represent the precision and recall of the predicted masks, respectively. 
    \end{tablenotes} 
    % }
    \end{table*}

For the validation phase, we enlist the help of three experienced pathologists to assess the effectiveness of the generated model. We use standard evaluation metrics, including detection precision, recall, and the F1-score, to evaluate our detection model.

\textbf{The generative model. }Unlike conventional generative networks, we introduce a self-supervised mask generation function to control the details of CD20 staining patch generation. Specific masks include those for H\&E cells, red blood cells, CD20 cells, and stained areas. These masks do not need to be labeled manually but are derived by automatically calculating the segmentation threshold using the multi-Otsu threshold method. During training, the generated virtual CD20 staining patches are used to generate the corresponding masks based on the previously determined thresholds. We use cross-entropy loss to compute the difference between the predicted mask and the actual mask, which is used to update the model parameters. This approach effectively stabilizes the color of the generated patches. Compared to generation without mask guidance, our method shows significantly better results, as shown in Fig.~\ref{m3}. Further, we use non-overlapping H\&E staining patches for feature transfer and then splice the generated results into a whole WSIs to compare the overall performance of different models, as shown in Fig.~\ref{mpj}.

In addition, we use the Fréchet Inception Distance (FID) to calculate the relative similarity between the generated data and the H\&E data, and assess the quality of the generative model by comparing the similarity distance between different generated data and the H\&E data. We believe that employing a standardized metric helps reduce variability in judgment across different doctors, leading to a more precise assessment of the generative model's effectiveness. Refer to Table~\ref{bfid} for the specific results.

\textbf{The detection model. }
For the detection model, YOLO, Mask R-CNN, Swin Transformer, and other models are commonly used for medical object detection. Considering the need to accurately detect tissues of smaller size and to minimize the number of parameters required by the model, we use the YOLOv5 detection model based on Darknet53 (an anchor-based, single-stage model structure that uses FPN and PAN to quickly and accurately localize smaller lesions) to validate the effectiveness of our approach. To demonstrate that one-to-one paired H\&E and CD20 staining patches effectively improve detection, we control the datasets structure for model training to avoid disturbances caused by randomly mixing H\&E and CD20 staining patches. Finally, one-to-one patches of H\&E and CD20 staining after concatenation are input into the model, and the final detection result is obtained. the visualization of each model's detection effects and the comparison of results shown in Fig.~\ref{m4} and Table~\ref{b2}.

\section{Conclusions} 
\label{conclusion}
In this study, we propose that the specificity of IHC can enhance the accuracy of TLS detection. Based on this assumption, we propose the Virtual Immunohistochemical Pathologic Analysis Network (VIPA-Net), an integrated framework that includes a Mask-Guided Transfer Module and an H\&E-based Virtual Staining Color Simulation TLS Detection Module. First, we use H\&E and CD20 staining patches and their masks to synthesize realistic IHC staining patches without requiring explicit label information. Then, we employ multiple staining patches for accurate TLS localization. Experimental results demonstrate that using multiple staining patches significantly improves detection performance.

\bibliographystyle{IEEEtran}
\bibliography{ijcai20}

\end{document}